\documentclass[prb,aps,amssymb,showpacs,twocolumn]{revtex4}

\usepackage{amsmath}
\usepackage{amssymb}
\usepackage{amsthm}
\usepackage{amsfonts}
\usepackage{algorithmic}
\usepackage{enumerate}
\usepackage{latexsym}
\usepackage[dvips]{graphicx}

\newcommand{\beq}{\begin{equation}}
\newcommand{\eneq}{\end{equation}}
\newcommand{\bed}{\begin{displaymath}}
\newcommand{\ened}{\end{displaymath}}
\input{epsf}

\begin{document}

\tolerance 10000


\title{On Gossamer Metals and Insulating Behavior}

\author { George Chapline$^\square$, Zaira Nazario$^\dagger$ and 
David I. Santiago$^{\dagger, \star}$ }

\affiliation{ $\square$ Physics and Advanced Technologies Directorate\\ 
           Lawrence Livermore National Laboratory, Livermore, CA 94550. \\
             $\dagger$ Department of Physics, Stanford University,
             Stanford, California 94305 \\ $\star$ Gravity Probe B Relativity 
             Mission, Stanford, California 94305}
\begin{abstract}

\begin{center}

\parbox{14cm}{ We extend the Gossamer technique recently proposed to
describe superconducting ground states to metallic ground states. The
gossamer metal in a single band model will describe a metallic phase
that becomes arbitrarily hard to differentiate from an insulator as
one turns the Coulomb correlations up. We were motivated by the phase
diagram of V$_2$O$_3$ and f-electron systems which have phase diagrams
in which a line of first order metal-insulator transition ends at a
critical point above which the two phases are indistinguishable. This
means that one can go continuously from the metal to the
``insulator'', suggesting that they might be the same phase. }

\end{center}
\end{abstract}


\maketitle

Vanadium sesquioxide V$_2$O$_3$, one of the so called Mott insulators,
has been widely studied because of its
very interesting phase diagram\cite{vanad}. Like most Mott
insulators, V$_2$O$_3$ is an antiferromagnet at low enough
temperatures. This may be a general phenomenon and all 
correlated insulators will order at a low enough temperature.

The phase diagram of V$_2$O$_3$ and other Mott insulators is such that the 
antiferromagnetism and insulation disappears with pressure at 
$T=0$\cite{vanad}. As one increases the temperature of the antiferromagnetic 
phase, the spin order melts and one is left with a spin disordered  phase that
is insulator-like in the sense that the resistivity is rising with
diminishing temperature. As one applies pressure to this insulating-like phase 
it undergoes a first order phase transition into a metal.

The interesting thing is that the transition line between the metal
and the insulator-like phase terminates at a critical point $(T_c,
P_c)$. Similar behavior is observed in f-electron compounds\cite{ce}. This
means that the metal and the insulating-like phase cannot be fundamentally 
different as one can go continuously from one into the other, analogous to
the continuous transformation of liquid into a gas above the critical 
temperature. We posit that the spin disordered insulator-like phase is a bad 
tenuous ``Gossamer'' metal rather than a true insulator. 

Recently Laughlin\cite{bob} and collaborators\cite{us} have advocated a 
similar Gossamer phenomenology as apt to describe the superconducting cuprates.
The main idea of the Gossamer superconductor is that there is a superconducting
pairing amplitude all the way to zero doping into the ``insulating'' regime. 
The Coulomb interactions decrease the superfluid density. The superconducting
d-wave spectrum will be there all the time, but the spectral weight
just becomes smaller and smaller as one goes to zero doping. The 
missing spectral weight goes into forming Hubbard-like bands. In the model 
introduced by Laughlin\cite{bob}, this spectral redistribution comes about
as result of partial Gutzwiller projection, which we call the Gossamer 
technique.

The Gossamer technique used for the superconductor should work 
for a metal in an analogous fashion, as has been recently 
proposed\cite{kollar}. Near full projection and near half-filling 
the density of states at the Fermi level collapses to zero. The missing 
spectral weight would go to forming Mott-Hubbard bands.  Such a ``Gossamer'' 
metal would beimpossible to tell apart from an insulator except at the lowest
possible temperatures in which, at least for the case of V$_2$O$_3$,
antiferromagnetic order intervenes and the material is truly
insulating. If antiferromagnetism does not intervene
at low temperatures, we would expect the resistivity to saturate to a large
but finite value at small temperatures absent localization effects.
We propose that something analogous to this is happening for the disordered 
insulating-like phase of some correlated electron systems.

We speculate that, while our original motivation for the present work
was V$_2$O$_3$, perhaps f-electron systems, which are charge transfer
insulators like the superconducting cuprates, are more apt to be
described as gossamer metals. In particular we draw attention to
f-electron collapse in the rare earth and actinide elements and their
compounds\cite{f}. This is usually a first order phase transition with a
volume change, and results from changes in composition or pressure. It
stands out as a characteristic feature of these materials. Indeed
understanding f-electron collapse may be of importance for
understanding the remarkable properties of these materials. A typical
example of f-electron collapse is the $\alpha \rightarrow \gamma$ phases
transition in elemental cerium, which occurs at room temperature at a
pressure near to 10 kbar\cite{ce}. This phase transition is thought to
be the result of an f-electron initially localized on a cerium ion
becoming delocalized as a result of increased hybridization with
conduction electrons.  Correlations between the f-electrons
are also thought to play an important role. Although f-electron
collapse is normally a first order phase transition, in the following
we would like to focus on the fact that since there is a critical
point the $\alpha$ and $\gamma$ phases of Ce can really be regarded as a single
Gossamer metallic phase.

The Gossamer technique is a bit more delicate to apply to a metallic Fermi sea 
ground state than to a superconducting BCS ground state. The reason is that 
the abundance of low energy degrees of freedom in the metal might make the
projection uncontrolled in the infrared leading to unphysical
results. Since the superconductor does not have such a plethora of low
energy degrees of freedom (irrespective of whether the gap has nodes
or not), the calculation has no infrared problems: it is regularized
by the superconducting order. We therefore will borrow heavily from
the calculations for the superconductor in previous
work\cite{bob,us}. After obtaining the results, we collapse the gap to
zero to study the physics of Gossamer metals.

We now proceed to reproduce and review the previous
results\cite{bob,us} and apply them to describe bad metallic
behavior. The Gossamer ground states are constructed by applying the
partial Gutzwiller ``projector''

\begin{equation}
\Pi_{\alpha_0} = \prod_j z^{(n_{j \uparrow} + n_{j \downarrow}) /
2}_{0}(1 - \alpha_0 n_{j \uparrow} n_{j \downarrow}) \; \; \; .
\end{equation}

\noindent $ 0 \le \alpha_0 < 1 $ is a measure of how effective the
projector is and in a real material it will be related to the Coulomb
repulsion. The factor of $z_0$, the quantum fugacity, in the projector
is the extra probability of having an electron at site $j$ {\it
chosen} to keep the total number of particles constant at
$(1-\delta)N$ after projecting. The fugacity is given by $z_0 =
(\sqrt{1 - \alpha(1-\delta^2)} - \delta)/ [(1-\alpha)(1-\delta)]$ with
$(1-\alpha_0)^2 = 1 - \alpha$\cite{bob,us}.

The Gossamer superconducting ground state is postulated to be $| \Psi
> = \Pi_\alpha \; | \Phi > \label{gs}$. Here $| \Phi >$ is the BCS
ground state:

\begin{equation}
|\Phi > = \prod_{\vec{k}} (u_{\vec{k}} + v_{\vec{k}}
c^{\dagger}_{\vec{k} \uparrow} c^{\dagger}_{-\vec{k} \downarrow})
|0> \; \; \; .
\end{equation}

\noindent where $u_{\vec{k}}$, $v_{\vec{k}}$ are the BCS coherence
factors given by $u_{\vec{k}}=$ $\sqrt{(E_{\vec{k}} +
\epsilon_{\vec{k}}- \mu)/2 E_{\vec{k}}}$ and $v_{\vec{k}}=$
$\sqrt{[E_{\vec{k}} - (\epsilon_{\vec{k}}- \mu) ]/ 2E_{\vec{k}}}$ with
dispersion $E_{\vec{k}}= \sqrt{(\epsilon_{\vec{k}} - \mu)^2+
\Delta^{2}_{\vec{k}}}$ where $\epsilon_{\vec{k}}$ is the kinetic
energy of the metal measured from the Fermi level, $\mu$ is the
chemical potential and $\Delta_{\vec{k}}$ is the superconducting
gap. Such projected ground states have been studied
before\cite{rand,zg,zg2,zg3}.

We never fully project ($\alpha_0 < 1$) in order for the partial
projector to have an inverse:

\begin{equation}
\Pi^{-1}_{\alpha} = \prod_j z^{-(n_{j \uparrow} + n_{j
\downarrow}) / 2}_{0}(1 + \beta_0 n_{j \uparrow} n_{j \downarrow})
\; \; \; ,
\end{equation}

\noindent with $\beta_0 = \alpha_0 / (1 - \alpha_0)$.  The Gossamer
ground state is the {\it exact} ground state of the Gossamer
Hamiltonian:

\begin{equation}
{\cal H} = \sum_{\vec{k} \sigma} E_{\vec{k}} B_{\vec{k}
\sigma}^\dagger B_{\vec{k} \sigma}, \;\; \; \; B_{\vec{k} \sigma}
|\Psi \rangle =0 . \label{gossham}
\end{equation}

\noindent where:

\begin{displaymath}
B_{\vec{k} \uparrow \{\downarrow\}} = \Pi_\alpha b_{\vec{k}
\uparrow \{\downarrow\}} \Pi_\alpha^{-1} = \frac{1}{\sqrt{N}}
\sum_j^N e^{i \vec{k} \cdot \vec{r}_j}
\end{displaymath}

\begin{equation}
\times \biggl[ z_0^{-1/2} u_{\vec{k}} (1 + \beta_0 n_{j \downarrow
\{\uparrow\}} ) c_{j \uparrow \{\downarrow\}} \pm z_0^{1/2}
v_{\vec{k}} (1 - \alpha_0 n_{j \uparrow \{\downarrow \}} ) c_{j
\downarrow \{\uparrow\}}^\dagger \biggr] 
\end{equation}

\noindent with $b_{\vec{k}\uparrow \{\downarrow\}} = u_{\vec{k}} c_{j
\uparrow \{\downarrow\}} \pm v_{\vec{k}} c_{j \downarrow
\{\uparrow\}}^\dagger$ the Bogoliubov quasiparticle operators which
annihilate the BCS ground state. The Gossamer Hamiltonian would be impossible 
to define if we used a full Gutzwiller projector.

The Gossamer ground state is adiabatically continuable to the BCS
ground state by continuously varying $\alpha_0$ to zero. Its uniqueness
follows from the uniqueness of the BCS ground state up to a
phase. Therefore the Gossamer superconductor describes the same phase
of matter as the BCS superconductor. In a similar fashion once we
collapse the gap to obtain the Gossamer metal ground state, it will be
adiabatically deformable to a regular Fermi sea metallic ground state
and, hence they will be the same zero temperature phase of matter.

In previous work\cite{bob} it was shown that the variational
wavefunction $| {\vec k} \sigma \! > = \Pi_\alpha b_{{\vec k}
\sigma}^\dagger | \Phi \! >$ represents an appropriate approximation
to the low energy quasiparticle excitations and that, remarkably,
their dispersion does not change under projection:

\begin{equation}
\frac{< \! {\vec k} \sigma | {\cal H} | {\vec k} \sigma \! >}
{ < \! {\vec k} \sigma | {\vec k} \sigma \! >}
= E_{\vec k} \; \frac{ < \! \Psi | \Psi \! >}
{< \! {\vec k} \sigma | {\vec k} \sigma  \! >}
\simeq E_k \; \; \; .
\end{equation}

\noindent We collapse the superconducting gap to find that the
dispersion in the the Gossamer metal is {\it unchanged} from the
dispersion in the regular metal. This might be a bit surprising as
usually correlation effects are thought to make the charge carriers
heavy\cite{heavy}. What is happening here is not that the carriers are
getting arbitrarily heavy as we approach the transition, but the
metallic band is becoming thinner and the missing spectral weight goes
to energies far from the Fermi sea to forming Hubbard bands. The
carriers are just as fast, there are just less of them which degrades
the conductivity.

From the matrix elements\cite{bob}

\begin{displaymath}
\frac{< \! \Phi | c_{j \uparrow} \Pi_\alpha^2
c_{j' \uparrow}^\dagger | \Phi \! >}
{< \! \Phi | \Pi_\alpha^2 | \Phi \! > } \times
\frac{ < \! \Phi | \Phi \! >}
{< \! \Phi | c_{j \uparrow} c_{j' \uparrow}^\dagger | \Phi \! >}
\end{displaymath}

\begin{equation}
\simeq \frac{4}{1 - \delta^2} (z z_0) \biggl[
\frac{1 + (1 - \alpha) z}{1 + 2 z + (1 - \alpha ) z^2} \biggr]^2 = 1
\end{equation}

\noindent
and

\begin{displaymath}
\frac{< \! \Phi | c_{j \uparrow}^\dagger \Pi_\alpha^2
c_{j' \uparrow} | \Phi \! >}
{< \! \Phi | \Pi_\alpha^2 | \Phi \! > } \times
\frac{ < \! \Phi | \Phi \! >}
{< \! \Phi | c_{j \uparrow}^\dagger c_{j' \uparrow} | \Phi \! >}
\end{displaymath}

\begin{equation}
\simeq \frac{4}{1 - \delta^2} (\frac{z}{z_0}) \biggl[
\frac{1 + z}{1 + 2 z + (1 - \alpha ) z^2} \biggr]^2 = 1
\; \; \; 
\end{equation}

\noindent the photoemission amplitudes were calculated for the Gossamer
superconductor. After collapsing the gap, for the Gossamer metal these are 
given by

\begin{displaymath}
\frac{ < \! - {\vec k} \downarrow | c_{{\vec k} \uparrow} |
\Psi \! > }
{\sqrt{< \! - {\vec k} \downarrow | - {\vec k} \downarrow > \;
< \! \Psi | \Psi \! > }} = 0
\; \; \; ,
\end{displaymath}
\begin{equation}
\frac{ < \! - {\vec k} \downarrow | c_{{\vec k} \uparrow}^\dagger |
\Psi \! > }
{\sqrt{< \! - {\vec k} \downarrow | - {\vec k} \downarrow > \;
< \! \Psi | \Psi \! > }} = g
\; \; \; ,
\end{equation}

\noindent with 

\begin{equation}
g^2 \simeq \frac{2 \alpha_0}{\alpha} \biggl\{ 1 - \frac{\alpha_0}{\alpha}
\biggl[ \frac{1 - \sqrt{1 - \alpha ( 1 - \delta^2)}}
{1 - \delta^2} \biggr] \biggr\}
\; \; \; .
\end{equation}

The suppression of photoemission amplitude is endemic of the smaller number of
metallic electrons whose number is diminished from that in the unprojected 
metal by a factor $g^2$ which goes to $2|\delta|/(1 + |\delta|)$ as $\alpha_0 
\rightarrow 1$. This is consistent with the previous result\cite{bob} that the 
superfluid density in the Gossamer superconductor goes like $g^2$ as there is a
sum rule for the superconductor making the density of conducting electrons 
equal to the superfluid density. For strong correlations the number of 
metallic electrons vanishes at half-filling. The material is metallic at all 
other dopings. Whether there is a full gap opened at zero doping is an 
interesting question. We have not yet been able to determine if the whole Fermi
surface is destroyed or if there are some Fermi points as the material is 
on the brink of becoming insulating and opening a gap. The continuous behavior 
with doping certainly argues for the latter.

We now proceed to analyze the specific form that the Gossamer metal Hamiltonian
takes. After some manipulation, we can bring the Hamiltonian into the form 

\begin{displaymath} \mathcal{H} = \sum_{\vec{k}} \frac{E_{\vec{k}}}{N} 
\sum_{i,j}^N e^{-i \vec{k}
(\vec{r}_i -\vec{r}_j)} \{z_0^{-1}  (1+\beta_0
n_{i\downarrow})(1+\beta_0 n_{j\downarrow}) c^\dagger_{i \uparrow}
c_{j \uparrow} 
\end{displaymath}

\begin{equation}
+ z_0^{-1}  (1+\beta_0 n_{i \uparrow})(1+\beta_0 n_{j
\uparrow}) c^\dagger_{i \downarrow} c_{j \downarrow} 
\end{equation}

\noindent We will see that the sums break up into a  chemical potential,
kinetic, as well as a Hubbard U terms. 

{\bf Kinetic Term in the Gossamer Metal}:
The off-site contributions of $\mathcal{H}$ give a hopping (kinetic)
term in the Hamiltonian. We can write $\mathcal{H}$ as a sum of
on-site and off-site contributions, ${\mathcal{H}} = {\mathcal{H}}_{on\; site} 
+ {\mathcal{H}}_{off\;site}$, where the two contributions read:

\begin{widetext}
\begin{equation} 
{\mathcal{H}}_{on\; site}= \sum_{\vec{k}} \frac{E_{\vec{k}}}{N} \sum_{j}^N  
\{z_0^{-1} (1+\beta_0 n_{j\downarrow})(1+\beta_0
n_{j\downarrow}) c^\dagger_{j \uparrow} c_{j \uparrow} + \{ \uparrow
\rightleftarrows \downarrow \}
\end{equation}

\begin{equation} 
{\mathcal{H}}_{off\;site} =\sum_{\vec{k}} \frac{E_{\vec{k}}}{N} 
\sum_{i \ne j}^N e^{-i \vec{k} (\vec{r}_i -\vec{r}_j)}
\times \{z_0^{-1}  (1+\beta_0 n_{i\downarrow})
(1+\beta_0 n_{j\downarrow}) c^\dagger_{i \uparrow}
c_{j \uparrow} + \{\uparrow \rightleftarrows \downarrow \}
\end{equation}
\end{widetext}

\noindent In this section we are interested only on the off-site
contribution. Without the partial Gutzwiller projection ($\alpha_0
=0$, $z_0 =1$) this term is just the kinetic term of a mean-field
Hamiltonian $\sum_{\vec{k} \sigma} (\epsilon_{\vec{k}} - \mu)
c^{\dagger}_{\vec{k} \sigma} c_{\vec{k} \sigma} $. However, the
partial projection induces complications which make
${\mathcal{H}}_{off\;site}$ very difficult to diagonalize
analitically. In order to get a rough estimate for the change in the
kinetic term due to partial projection, we will make the mean field
approximation $\langle n_{i\uparrow} \rangle = \langle n_{i
\downarrow} \rangle =\frac{1}{2}(1-\delta)$, and we will replace the number 
operators with this average. The term then becomes:

\begin{equation} 
{\mathcal{H}}_{off\;site}
=\sum_{\vec{k} \sigma} \frac{E_{\vec{k}}}{N} \sum_{i \ne j}^N
e^{-i \vec{k} (\vec{r}_i -\vec{r}_j)} z_0^{-1} (1+\frac{\beta_0(1-\delta)}{2}
)^2 c^\dagger_{i \sigma} c_{j \sigma}
\end{equation}

 \noindent This term finally is: 

\beq {\mathcal{H}}_{off\;site} = z_0^{-1} (1+
\frac{\beta_0(1-\delta)}{2})^2\sum_{\vec{k} \sigma}
(\epsilon_{\vec{k}} - \mu) c^\dagger_{\vec{k} \sigma} c_{\vec{k}
\sigma} \eneq 

\noindent Therefore the effect of the partial projection on the
kinetic term in the Gossamer Hamiltonian is an overall constant, a
renormalization of the energy by

\begin{equation}
\frac{[2 - \alpha_0(1+\delta)]^2(1 - \delta)}{4 \left(\sqrt{1 - \alpha(1- 
\delta^2)} - \delta \right)} \; ,
\end{equation}

\noindent with $1-\alpha = (1-\alpha_0)^2$. As we can see, the constant blows 
up as one approaches full projection $\alpha_0 \rightarrow 1$. This does
not lead to any complications since as we will soon see, the other
terms blow up in an exactly similar fashion leaving their physically
relevant ratio finite. When unprojected the renormalization factor is
1 as it should be.

{\bf Hubbard-U Term in Gossamer Metal}:
The Hubbard-U term will arise out of the on-site contribution,
${\mathcal{H}}_{on \; site}$  which was explicitly written down in the
previous section. One can write the terms as: 

\begin{displaymath} 
{\mathcal{H}}_{on \; site} = \sum_{\vec{k}} \frac{E_{\vec{k}}}{N} 
\sum_j^N \{ z_0^{-1}  (n_{j \uparrow} 
+ n_{j \downarrow}) +
\end{displaymath}

\beq 
+z_0^{-1} (4\beta_0 + 2 \beta_0^2)n_{j \uparrow}  
n_{j\downarrow}\} 
\eneq 

\noindent The first term is a chemical potential, and the second term
is the Hubbard-U. We will concern ourselves only with this last
term. For the factor containing $\beta_0$ we obtain:

\beq 
z_0^{-1} (4\beta_0 + 2\beta_0^2)  
=  \frac{2\alpha_0(2 - \alpha_0)
(1 - \delta)}{ \sqrt{1 - \alpha(1- \delta^2)} 
- \delta } \; 
\eneq 

\noindent and therefore the Gossamer metal will have a Hubbard U term
of the form $\sum_j^N U n_{j \uparrow}  n_{j \downarrow}$ where, at 
half-filling 

\beq 
U= \sum_{\vec{k}} \frac{E_{\vec{k}}}{N} 
\frac{2\alpha_0(2 - \alpha_0)(1 - \delta)}
{ \sqrt{1 - \alpha(1- \delta^2)} - \delta }  
\eneq

\noindent As we can see, at full projection $\alpha_0 \rightarrow 1$
this term blows up as well, but the ratio between the hopping
amplitude computed in the previous section and the Hubbard U remains
finite and is of order 1 near full projection. When unprojected, the U
term is zero as it should be.

The existence of the growing Hubbard U term means that as we go to
half-filling and full projection magnetic correlations will get
enhanced leading to a diverging magnetic susceptibility in the exact
same way as for the Gossamer superconductor\cite{us} after we collapse
the gap. The spectral weight will consist of Mott-Hubbard bands at
high energies with the chemical potential pinned at midgap at an ever
fainter band from which the Gossamer quasiparticles are excited with
a dispersion unchanged from the noninteracting metal. 

As mentioned before, this unmodified dispersion is different behavior from what
was found in previous investigations of the Hubbard model with Gutzwiller 
projectors by Brinkman and Rice\cite{heavy}. The difference stems from the fact
that the projection technique used in the previous work since it did not 
include a
fugacity factor, $z_0$, to conserve particle number. This leads to an 
arbitrarily large $U$ and finite $t$ leading to a diverging carrier mass as one
goes to half-filling. In fact, a recent proposal for a Gossamer metal ground  
state\cite{kollar} with projectors that do not conserve the number of particles
finds the exact same behavior as in the Brinkman-Rice work\cite{heavy}. In this
note we make sure to conserve particle number, the ratio $U/t$ remains finite
and below the critical value for true insulation, and the carriers cannot 
become arbitrarily heavy.

In the present work we extended the Gossamer technique originally
developed for the superconducting cuprates to metallic ground
states. The Gossamer metal describes strongly correlated bad metal
behavior that is very hard to distinguish from true insulating
behavior, the implication being that the magnetically disordered
insulating-like phases in some systems might really be a Gossamer metal
with very much degraded conductivity. The degraded conductivity
arrises from a depletion of spectral weight of metallic electrons and
not by an ever growing effective mass of the carriers. We studied this
in a single band model. The single band model might not be an apt
description of the f-electron system, which was one of the motivating
players for the present work. Despite this, we believe that the
general features we have described are general and should survive in
such systems.

{\bf Acknowledgements}: We thank Bob Laughlin for valuable
discussions and suggestions. We thank B. A. Bernevig and D. Giuliano for 
discussions.

\end{document}